\documentclass{article}
\usepackage{ijcai21}

\usepackage{times}
\usepackage{soul}
\usepackage{url}
\usepackage[hidelinks]{hyperref}
\usepackage[utf8]{inputenc}
\usepackage[small]{caption}
\usepackage{graphicx}
\usepackage{amsmath}
\usepackage{amsthm}
\usepackage{booktabs}
\usepackage{algorithm}
\usepackage{algorithmic}
\title{Survey of Parallel A* in Rust}

\author{
Brett Fazio$^1$\and
Ellie Kozlowski$^1$\and
Dylan Ochoa$^1$\and
Blake Robertson$^1$\and
Idel Martinez$^1$\\
\affiliations
$^1$University of Central Florida\\
\emails
\{brettfazio, elliekozlowski, dochoa23, blakewrobertson, idel\_martinez\}@knights.ucf.edu,
}

\begin{document}

\maketitle

\begin{abstract}
A* is one of the most popular Best First Search (BFS) techniques for graphs. It combines the cost-based search of Breadth First Search with a computed heuristic for each node to attempt to locate the goal path faster than traditional Breadth First Search or Depth First Search techniques. However, A* is a sequential algorithm. The standard implementation only runs in one thread. There are a few attempts to get A* to leverage multiple threads. Centralized (SPA*) and Decentralized (DPA*, HDA*) methods are the most standard attempts, with the most unique and modern method being massively-parallel A* (MPA* or GA*). We will attempt an implementation of each in Rust to determine if there is a performance boost, and which one has the best performance. 
\end{abstract}

\section{Introduction}

This paper surveys the state of the art of parallel A* and Best First Search algorithms. We implement several approaches and empirically evaluate the performance of each. Our goal is to survey each of these methods in Rust \cite{rust} - a relatively new language without Parallel A* implementations but with all the tools to create them.

We implemented KPBFS, DPA*, and HDA* in Rust. We overview the research of each of these methods, detail our implementations, and then go into detail on analyzing the results of our experiments. We test each of our implementations using multiple heuristics on 1, 2, 4, 8, and 16 threads.



\section{Background \& Related Work}

\subsection{A*}

A* was first proposed as a solution to finding a minimal cost path through a graph in 1968 \cite{4082128}. Since the details of the algorithm are well known, we will not go into too much detail going over it, however we will present again the concepts utilized in the parallel implementations.

A* computes the minimum-cost path in a graph from the $START$ node to the $END$ node. The core components of A* are in the ordering of the nodes. As nodes are added to the open set to process, they are ordered according to $ f(n) $ where $ f(n) = g(n) + h(n) $. Here, $ g(n) $ is the cost that it takes to get from $START$ to $n$, and $ h(n) $ is the computed heuristic cost. In our setup each node will have Cartesian metadata ($x$ and $y$ coordinates) on which we will compute Manhattan distance between the current node and $END$. The Manhattan distance would be guaranteed to be an admissible heuristic as in our graph setup, movement is only allowed in one of four cardinal directions with some directions being blocked off, resulting in a cost of $c$ from $n$ to $END$, where $c \geq h(n)$.

In addition to Manhattan distance, we may also explore additional heuristics to see how heuristic quality, accuracy, and complexity affects different runtimes.

\subsection{Classification of Methods}

The two different approaches to implementing parallelized A* are centralized and decentralized. The centralized approach ``requires synchronization on every node, expansion, and generation" \cite{survey_p_astar}. The decentralized approach introduces the problem of work distribution. In the literature, the solutions to this problem are hash-based work distribution, structured abstraction, and randomized strategies. The hash-based work distribution is widely explored and divided into synchronous and asynchronous communication. The different implementations use different feature generation methods.

\subsection{Centralized A*}

One of the first methods of parallel A* proposed is Centralized Parallel A* (CPA*) \cite{cen_decen}. Fukunaga et al. proposes Simple Parallel A* (SPA*) as the simplest way to do CPA* \cite{survey_p_astar}. SPA* is implemented in a quite simple derivative of normal A*. By placing a lock on the $OPEN$ list, and making both the $OPEN$ and $CLOSED$ lists shared memory, SPA* just deques nodes across multiple threads until the goal is reached. The problem is that since a lock is used and threads will likely be waiting on one another, SPA* has worse runtimes than single threaded A* (which is why we are not pursuing the implementation of SPA*) \cite{survey_p_astar}. Additionally, given that we are pursuing an implementation of KPBFS (\ref{kpbfs_impl}), the two implementations are incredibily similar and KPBFS has use cases unlike CPA*, which is why we see providing a KPBFS implementation as sufficient.

\subsection{K-Best First Search}
\label{kpbfs_impl}

The K-Best First Search algorithm (KBFS) is an algorithm that expands the top $K$ nodes from the $OPEN$ list. With this it provides better (albeit approximate) results when given inadmissible heuristics \cite{k_bfs}. KBFS most accurately models real world situations where large errors in heuristic calculation can occur. 

An extension of KBFS, entitled KPBFS, \textit{P} meaning \textit{P}arallel, was introduced later in an attempt to exploit the increased performance of running many threads at once \cite{kpbfs}. As it is a derivative of KBFS, KPBFS's goal is to find approximate solutions to inadmissible, or sub-optimal, BFS problems. KPBFS has its best performance when the heuristics are computationally expensive, thus utilizing the threads to a greater degree and guaranteeing a coarse-grained approach \cite{kpbfs}. While performance increases were not in the double digits, a better performance was seen when the number of threads was increased (even when using more threads than cores) \cite{survey_p_astar}.

\subsection{Decentralized A*}
A decentralized version of parallel A* reduces the contention caused by the centralized parallel A* \cite{cen_decen}. In particular, it takes advantage of giving each processor its own local $OPEN$ and $CLOSED$ lists. We will be investigating some of the advantages and disadvantages of decentralized parallel A*, as well as some communication methods used to improve its performance \cite{2017arXiv170805296F}.  One option thread communication method is to have a $BUFFER$ list stored on each thread but allow communication across all threads.  This helps threads randomly assign and distribute the work when a node looks at its neighboring nodes.  This we hope will also lead to better performance as more threads in the system will be aware of the optimal solution quicker instead of getting settled on one particular non-optimal path.

HDA* is an extension of DPA* that is supposed to provide greater performance via a hash-based distribution of work in which each processor comes to own a portion of the search space. \cite{survey_p_astar} We hope an implementation of HDA* will improve upon an implementation of DPA*.

\subsection{Massively Parallel A*}

Massively Parallel A* research focuses on how one can use the power
of a GPU to make a graph algorithm such as A* increasingly faster. Proposed in \cite{survey_p_mpastar} the first step to exploiting the parallel processing nature of a GPU is the ability to used parallel data structures and algorithms because most common ones are written sequentially in nature to run on a CPU.  They propose an algorithm GA*.  

This algorithm at first glance is not much different than DPA* but differs in a few ways.  While DPA* utilized OPEN and CLOSED Lists to maintain the state of the nodes, GA* to exploit the nature of a GPU utilizes parallel queues.  GPU power is also shown because of just using 1 queue similar to an OPEN list in DPA*, GA* utilizes thousands of queues in parallel to get the best performance.

\subsection{Motivation}

With all these cutting edge methods of creating parallel A*, the results in the simple case are a slight performance increase \cite{survey_p_astar}. However, when the problem becomes more complex or large, certain methods start to shine.

With KPBFS, situations where the heuristic is costly or where it is inadmissible result in KPBFS having performance benefits over KBFS and A* \cite{kpbfs}.

With DPA*, if the graph is especially large, that is $V$ the set of open nodes, grows significantly as the search progresses, DPA* is shown to outperform A* \cite{cen_decen}.

\section{Methodology Overview}

Concurrent programming in Rust is achieved through the \texttt{std::thread} module, which gives us access to native threads. The \texttt{spawn} method, which starts a thread, returns a \texttt{JoinHandle} struct,  giving the main thread control of the other threads for blocking, unblocking, and waiting for termination; that is, when the algorithm finishes on success or failure.

\label{MPSC}
For communication, the standard module allows for the use of channels, atomic shared variables, atomic reference counters (\texttt{Arc}s), locks, and barriers as part of the \texttt{std::sync} module. Shared variables can be implemented with a combination of locks (\texttt{Mutex}) and \texttt{Arc}s which wrap any data structure or variable, or through atomic types, which are only implemented for unsigned and signed integers and booleans. However, this introduces additional overhead as the contention for a variable increases when the number of threads present increases. For this reason, Rust provides asynchronous channels to send data between threads, allowing for communication without blocking, as part of the \textbf{m}ulti-\textbf{p}roducer \textbf{s}ingle-\textbf{c}onsumer (mpsc) strategy - meaning that multiple threads can send messages to the same receiver but only one thread is associated with the receiver. Here, threads can send data to each other, which is added to the receiver's bounded or unbounded buffer, so that it can be received at a later time. 

Channels, although useful, can cause issues in a parallelized algorithm when threads need a checkpoint to be forced to wait for all other threads to receive data, since the buffer may get too large. For this, Rust allows the use of barriers (as part of ``thread synchronization") which makes all threads wait at a given point for the specified number of threads to catch up, thus forcing all threads to read the most updated data. A flaw, however, is found when a thread exits on success, but is not able to notify the barrier (given its implementation), forcing all threads to reach the same checkpoint and deadlock since the exited threads will never reach it. For this reason, we implemented our own version of a dynamic barrier.

\section{Centralized A*}

\subsection{SPA*}

As discussed previously in Fukunaga et al. \cite{survey_p_astar}, the most standard CPA* approach, SPA*, has been shown to have worse runtimes than a single-threaded, Best First Search - thus we did not see it as beneficial to provide an implementation in Rust.

\subsection{K-Parallel Best First Search}

KPBFS was the first parallel A* that we created. Following closely the work of Vidal et al. we used their outline to create our implementation of KPBFS. \cite{kpbfs} No psuedocode was provided so some interpretation of the paper had to be done. 

At the implementation level, our KPBFS solution in Rust is the easiest to understand as it resembles the traditional A* algorithm the most. KPBFS uses locks in the original implementation so we utilize them here as well. 

Our implementation works by starting up a series of search functions each in their own thread - they share Mutex $HashMap$ of explored states and a Mutex $PriorityQueue$ of the elements to be explored next. Within the search function a classic A* is implemented with the main difference just being that any operations that access shared memory (popping off the next state, modifying the closed list, etc) must obtain a lock first. The program terminates when the shared queue is empty - as at that point there are no longer any relevant states left to explore.

\section{Decentralized A*}
As stated in section II, decentralized parallel A* takes advantage of giving each processor its own $OPEN$ and $CLOSED$ lists. The algorithm would begin by having a single processor delegating nodes to each available processor. Then, each processor runs its own localized A* search using its local $OPEN$ list. This removes the concurrency overhead that was generated by the centralized parallel A* implementation. With no communication between the processors, there would be a lot of repeated computation when expanding nodes. For this reason, we discuss two communication strategies and their implementation in Rust, with the goal of reducing redundant expansions.

 \subsection*{The Black Board Communication Strategy}
There are a plethora of communication methods available for us to use to assist in the load balancing for the decentralized parallel A* algorithm. One such method includes The Black Board Communication Strategy, in which we have a shared ``black board" among all processors. The processor looks at its local $OPEN$ list; if its best node is much better than the best node on the black board (meaning it has a much lower $f$-value relative to the threshold we set), we will give the blackboard some of the ``good" nodes from our local $OPEN$ list. If the processor's best node is significantly worse than the best node on the black board, we will move some of the best nodes on the black board to the local $OPEN$ list. The idea is that we want to always be expanding the nodes with the lowest $f$-values\cite{cen_decen}.

\subsection*{The Random Communication Strategy}
The idea for this communication method is when the successor nodes are generated for a specific processor, we distribute these nodes randomly to other available processors' $OPEN$ lists. That way, other processors get a partition of the ``good part" of the search space. An issue with this approach is that duplicate nodes could end up being expanded if they exist in one processor's $CLOSED$ list but go to a different processor's $OPEN$ list. We can implement a search for duplicates, but this would lead to a lot of search overhead.

\subsection*{Implementation Details}
\label{dyn_barrier}
To achieve these strategies, threads communicate through channels, as shown in Section III, which permits us to send the data to the hash selected nodes, who then add the received data to their respective $OPEN$ list. To avoid having threads reading an empty buffer, we introduced a barrier that synchronizes all threads. An issue occurs, as discussed earlier, when one or more threads exit early, causing the other threads to enter a deadlock when executing \texttt{barrier.wait()}. Because the barrier does not detect threads that exited and the count for threads waiting cannot be decremented, we introduce our implementation of a dynamic barrier that can correctly synchronize threads at a checkpoint without deadlock. Its design is modeled after the current barrier, but waiting is implemented through channels, allowing threads to signal when they exited early, thus preventing the deadlock.

A thread terminates either when its amount of sent messages is equal to the amount of received messages or when the early exit flag is set to true. This early exit flag is set to true when the thread finds a better path to the goal node than the one that previously existed.

\section{Hash Distributed A*}

Hash Distributed A* is a decentralized implementation that utilizes hash-based work distribution. This method was a reopened investigation into hash-based work distribution for A* by Kishimoto, Fukunaga, and Botea \cite{survey_p_astar}. The hash function is the key to creating any substantial parallel speedups in this implementation. In HDA*, each thread is given a part of the search space by using the hash function. Both the open and closed lists use a distributed data structure. The way the partitioning works is the thread waits until it has received a message (a new state to potentially expand) and then checks if the state (from the message) is in the closed list. If not, the new state is inserted into the open list. Then, if there are currently no messages, the thread will select the highest priority state from its open list and expand it. Expanding generates new states; for each new state, a hash key is computed based on the state’s representation. The thread then sends a message to some other thread who “owns” the new state according to its hash. 

Rather than having the open and closed lists shared amongst the threads in a way that requires synchronization, HDA* does not “explicitly” share these lists \cite{survey_p_astar}. HDA* uses message passing, with state-packing as an optimization, to communicate amongst threads. 

\subsection{Implementation Details}

For our implementation of HDA* we begin by expanding the start node. After this each thread executes the following steps until the optimal solution is found. First, the thread checks if any states have been received in its message queue. If there are any, then it will check to see if the state is in the closed list which determines whether the state is a duplicate or if it should proceed to be inserted into the open list. Second, if there are no messages in the queue then the thread will grab the highest priority state from the open list and expand it \cite{survey_p_astar}.

One of the biggest differences between this implementation and our DPA* implementation is that here we perform all of the checks previously done in the for loop controlled by the buffer list right when we receive a node. Then once we are done receiving, we expand the root of the open list. In short, there is no buffer list and some of the logic is reordered.

\section{Massively Parallel A*}

Massively Parallel A* is a new concept that primarily focuses on using GPU computing power to increase speed of existing A* algorithms.  GPUs optimize for parallel processing unlike CPUs that are designed to work better for sequential processes.  One of the bigger bottlenecks in A* is computing the heuristic function.  This can be greatly reduced by using GPUs since we can exploit the design of GPUs by paralleling the computation.  Another problem is the fact that when expanding nodes they often only have a very small finite number of states to process while most GPUs have thousands of cores.  These states are often processed using a priority queue which are often sequential in nature and even the lock-free parallel priority queues do not run efficiently on most GPU architectures.  Zhou and Zeng propose a new way of GA* which optimizes the A* algorithm to work on GPUs in the intended manner.  The GA* algorithm uses K priority queues to expand the nodes in the graph.  The larger K is the better we can exploit the parallel nature of GPUs but can not increase infinitely because of the overhead it would create.  Each priority queue would then extract states.  These states would then be expanded and duplicated in order to then compute their corresponding heuristic values.  Theses values would then be pushed back onto the the queues.  The GA* algorithm is then terminated when the queues are empty.  This algorithm showed great speed up over A* algorithms on CPUs and even other MPA* algorithms on GPUs. \cite{survey_p_mpastar}       

\section{Evaluation}

\subsection{Timing}

Timing is crucial for evaluating the performance of each implemented method. For timing we will utilize the Rust standard library, superficially \textit{std::time} as it suits our needs perfectly without reinventing the wheel \cite{rust}. We will also utilize Rust's \textit{bench} functionality to create workflows that automatically benchmark our implementations with various flags for thread count and heuristic type.

\textit{bench} provides us with millisecond levels of detail, outliers, and regressions so it is more than powerful enough for our needs. Additionally by programmatically creating a number of test scenarios we can run them from the command line which means we can test without having to make minor changes in our code. 

The output produced by \textit{bench} also runs each function a number of times and producing an average ensuring that our metrics are not just one-off runs but rather representative of the performance as a whole. 

\subsection{Graph Generation}
In order to test the correctness and runtime of our parallel A* implementations, we will be generating a set of test data. A test case will include a number $n$, representing the length and width of the square grid. The grid will be made up of the characters ‘.’, ‘S’, ‘E’, and ‘W’ which represent a node, the start node, the end node, and a wall, respectively. The goal is to reach the ‘E’ node starting at the ‘S’ node by only taking the open paths, ‘.’, and avoiding the walls, ‘W’. The test cases consist of bounds where $n = 10, 100, 1000$ and it is guaranteed that there will be a valid path from ‘S’ to ‘E’. We randomly generated the walls so that our A* implementations will be tested against more complicated traversals than a simple breadth first search. The nodes have a 20\% chance of being a wall, and the start and end nodes are also randomly generated.

\subsection{Heuristic}

\subsubsection{Admissibility}

A* is said to be admissible when the heuristic used is also admissible; that is, the heuristic never overestimates the cost of reaching a goal \cite{4082128}. KBFS differs from classical A* in that it is designed to work with inadmissible heuristics and provide approximate solutions in said cases \cite{k_bfs}. As such, with our KPBFS solution, we will simulate inadmissible heuristics by adding random noise to our euclidean distance calculation. 

\subsubsection{Cost}

As stated in KPBFS \cite{kpbfs}, the greatest parallelized returns are seen when the heuristic is computationally expensive. When cheap (in a computational context) heuristics are used, Vidal et al. states that less time is spent in a truly parallel setting and more time dequeing and expanding nodes \cite{kpbfs}. As part of our evaluation, we will use computationally cheap heuristics with each implemented method as well as a simulated expensive method for each. This will ensure that if a certain parallel method works well for expensive heuristics or poorly for cheap ones we will have accurate data points for both.

Our cheap heuristic will just be the computation of the 2D Euclidean distance (given the $(x, y)$ metadata for each node). Our expensive heuristic will be a simulated complexity, sleeping the thread for some pseudo-random amount of time greater than the time it takes to compute Euclidean distance but still less than a few seconds.

\subsubsection{Evaluation}

Given the details of admissibility and cost, we have decided to create 5 heuristic types to benchmark our functions. Each of the 5 types are outlined and explained below.

\begin{itemize}
    \item Euclidean - computes the Euclidean distance between two points on the grid defined by $\sqrt{(x_1-x_2)^2 + (y_1-y_2)^2}$. 
    \item Manhattan - computes the Manhattan distance between two points on the grid defined by $|x_1-x_2| + |y_1-y_2|$.
    \item Inadmissible - computes the Euclidean distance as defined above and then adds a small random delta to make the heuristic and over-estimation and thus inadmissible.
    \item Expensive - computes the Euclidean distance as defined above but the method takes longer to execute by sleeping the thread for a small number of milliseconds thus simulating a computationally expensive heuristic as outlined in different reference papers. \cite{kpbfs}
    \item ExpensiveInadmissible - a combination of the Expensive and Inadmissible methods above. It adds a small random delta to make the Euclidean distance and over-estimation (and thus inadmissible) as well as waits the function for a small amount of time.
\end{itemize}

Our testing mostly made use of the Euclidean heuristic as a baseline and the Expensive heuristic to simulate the conditions specified by the various reference papers. The Inadmissible heuristic was used to test the inadmissibility strength of the KPBFS algorithm compared to other implementations.

\section{Results}

\subsection{KPBFS}

\subsubsection{Background}

As Vidal et al. states in their KPBFS paper - they see the most performance benefit when the heuristic is costly as the algorithm itself is coarse-grained and less benefits when the heuristic is  computationally cheap. \cite{kpbfs} Both of these findings were reflected in the results of our implementation.

\subsubsection{Cheap Heuristic}

In this run the Euclidean heuristic is used as defined previously. A search space of 5000*5000 nodes is used as it was found to be large enough for the algorithm to run in more a few milliseconds while still not taking exceedingly long. Other size search spaces showed the same trend.

\begin{table}[htb]
  \centering
\begin{tabular}{|c|c|}
  \hline
  Thread Count & Run Time (Seconds) \\
  \hline
  1 & 0.41948 \\ 
  \hline
  2 & 0.42938 \\ 
    \hline
  4 & 0.50935 \\ 
    \hline
  8 & 0.81404 \\ 
    \hline
  16 & 1.3249 \\ 
  \hline
\end{tabular}
\caption{KPBFS results with Euclidean heuristic on a 25,000,000 node search space}
\label{table:kpbfs_euc}
\end{table}

As shown in \ref{table:kpbfs_euc} since this is a coarse-grained solution with high amounts of contention increasing the thread count decreases the runtime. This is not to say that the algorithm is completely useless - in fact this was outlined in \cite{kpbfs} as the reality of the algorithm and the places it shines are with expensive heuristics and inadmissible heuristics.

Looking at the results in finer detail the regression isn't too significant when going from 1 to 2 to 4 threads - however when looking at 8 and 16 they are 2 times and 3 times the originally runtime respectively which is a major regression. This would be worrying if this were the intended use case for KPBFS, but as explained above it is not.

\subsubsection{Expensive Heuristic}

In this run the expensive heuristic took 1 millisecond to compute. A smaller 10,000 node search space was used as the increased cost of the heuristic would result in extremely large run-time for larger search spaces. Other size search spaces showed the same trend.

\begin{table}[htb]
  \centering
\begin{tabular}{|c|c|}
  \hline
  Thread Count & Run Time (Seconds) \\
  \hline
  1 & 4.6682 \\ 
  \hline
  2 & 2.3360 \\ 
    \hline
  4 & 1.2112 \\ 
    \hline
  8 & 0.59077 \\ 
    \hline
  16 & 0.29615 \\ 
  \hline
\end{tabular}
\caption{KPBFS results with Expensive heuristic on a 10,000 node search space}
\label{table:kpbfs_exp}
\end{table}

The results of \ref{table:kpbfs_exp} show the strength of the KPBFS implementation - real world situations. In situations where the heuristic is more computationally complex than just a square root function (like with self driving car scenarios) KPBFS would provide noticeable performance improvements. By doubling the thread count a roughly 50\% cut in runtime is seen without any slowing - moving from 1 to 2 was a 50.04\% cut and moving from 8 to 16 was a 50.129\% cut. This test was obviously more of a simulation (of a more complex scenario) than a real world scenario but quantitatively confirms the results of Vidal et al. \cite{kpbfs}

\subsubsection{Inadmissible Heuristic}

KPBFS's goal with inadmissibility is that traditional A* algorithms will result in a goal being found with an incorrect cost when an inadmissible heuristic is used while KPBFS will still attempt an approximate answer. \cite{kpbfs} 

Using KPBFS with an inadmissible heuristic yields an approximate - albeit higher than correct - cost. Our hypothesis was that KPBFS should provide a better approximation than a more traditional A* (or DPA* or HDA*). This wasn't the case in our testing, the approximate answer on a 1000x1000 node graph was roughly the same between the various implementations with HDA* typically being the closest to the correct answer. We theorize this to be because of HDA's distribution of nodes so a more proper route was still explored. Even though KPBFS approximated similarly to HDA and DPA we additionally theorize its due to the multi-threaded nature of all the algorithms (they share the same advantages) where a comparison to a vanilla single threaded would result in KPBFS having a greater advantage. 

\subsection{DPA*}

\subsubsection{Background}

In Fukunaga et al.'s survey of Parallel A* they stated that DPA* provided significantly less overhead than centralized implementations however did not state directly that greater performance than single threaded A* could be directly achieved citing new problems that arise in distributed contexts such as balancing, re-exploration, and work stealing - and since this is still an actively researched topic there is no universal solution to all those problems. \cite{survey_p_astar}

Kumar et al., in their proposal of SPA* and DPA*, discussed the implementation as well as their initial findings. Extreme parameters were used, including a $2^{65}$ search space (significantly reduced by an admissible heuristic) and a 256-core CPU. \cite{cen_decen} The declared that in scenarios where the set of node expands very rapidly the decentralized parallel implementation out performs a single threaded implementation. 

The exact settings Kumar et al. used couldn't be replicated (namely the extreme search space and number of CPU cores), all testing was done on a 4-core CPU and specified search space sizes going up to $25,000,000$.

\subsubsection{Cheap Heuristic (Initial Run)}

Initial testing was done using the cheap heuristic as no major mentions of using computationally expensive heuristics were made in related works. Euclidean distance was also used to keep consistency with KPBFS results.

\begin{table}[htb]
  \centering
\begin{tabular}{|c|c|}
  \hline
  Thread Count & Run Time (Seconds) \\
  \hline
  1 & 0.084029 \\ 
  \hline
  2 & 0.22357 \\ 
    \hline
  4 & 0.2279 \\ 
    \hline
\end{tabular}
\caption{Initial DPA* results with Euclidean heuristic on a 1,000,000 node search space}
\label{table:dpa_euc}
\end{table}

The initial DPA* results we got from Kumar et al.'s outline of DPA* are above. \cite{survey_p_astar} A euclidean heuristic and a 1000 by 1000 graph are used. 

You may notice that only 1, 2, and 4 threads are shown here. This is because in the reference DPA* going beyond 4 threads on our machines resulted in our \textit{Buffer}s becoming backlogged and the program to hang with about a $30\%$ rate. We theorize this to be because of the differences in our machines as well as the space size. Another possible cause is that while a euclidean heuristic is admissible it is not 100\% accurate as all our implementations move in the 4 cardinal directions (manhattan movement) - thus the Manhattan heuristic would be more accurate but \textbf{both} would be admissible.

Looking at the results - we can see that 1 thread gives very good performance compared to the 2 and 4 threads - so some contention occurs when more threads are added. However, looking at 2 and 4 threads we can see somewhat promising results that no additional contention is added - over a few hundred runs of both 2 and 4 threads on DPA with the euclidean heuristic both are $~0.22$.

The below table provides the same test as above (DPA*, 1,000,000 node graph) but using the Manhattan heuristic as it provides a more accurate representation of the movement.

\begin{table}[htb]
  \centering
\begin{tabular}{|c|c|}
  \hline
  Thread Count & Run Time (Seconds) \\
  \hline
  1 & 0.050944 \\ 
  \hline
  2 & 0.12353 \\ 
    \hline
  4 & 0.11673 \\ 
    \hline
  8 & 0.129 \\ 
    \hline
\end{tabular}
\caption{Initial DPA* results with Manhattan heuristic on a 1,000,000 node search space}
\label{table:dpa_man}
\end{table}

We can see that the usage of the more accurate Manhattan heuristic does give noticeable benefits over the less accurate, but still admissible, Euclidean heuristic. At each thread level the speed is roughly double the speed of the Euclidean test, and in addition to this the 8 thread test is more stable (most likely due to the higher accuracy of the heuristic and not exploring bad states) and thus an averaged sample was able to be created.

\subsubsection{Cheap Heuristic (No Locks)}

This section just serves to provide some updated DPA* results - in our initial runs we used a \textit{Mutex} to keep track of the best solution (what we call the \textit{Incumbent}). We were able to change this to an atomic value with a compare and set loop as it made more sense with the situation and the results subsequently improved as you can see from the above below.

\begin{table}[htb]
  \centering
\begin{tabular}{|c|c|}
  \hline
  Thread Count & Run Time (Seconds) \\
  \hline
  1 & 0.051331 \\ 
  \hline
  2 & 0.078279 \\ 
    \hline
  4 & 0.072629 \\ 
    \hline
  8 & 0.092850 \\ 
    \hline
\end{tabular}
\caption{No lock updated DPA* results with Manhattan heuristic on a 1,000,000 node search space}
\label{table:dpa_man_no_lock}
\end{table}

As you can see from the table the results are significantly better when multiple threads are used - it is about a 2x speed up from the solution that uses a \textit{Mutex}.

\subsubsection{Cheap Heuristic (Updated Run)}
\label{update}
In the initial run we were unable to get 8 or 16 threads due to a deadlock issue that occurred. We discovered it to be because of the crossbeam library, specifically the \texttt{unbounded} channels that allow for them to be MPMC (multiple-producer, multiple-consumer), which were utilized in DPA* and HDA* for a simplification of wait and lock-free cross-thread sending and receiving of nodes. This allowed for the cloning of transmitters and receivers, which permitted each thread to have transmitters to all other threads. Following the pattern for MPSC (multiple-producer, single-consumer) from section \ref{MPSC}, it was believed that as soon as a thread exits, causing its receiver to go out of of scope, the receiver would be dropped (killed). However, the library's implementation of unbounded channels does not keep references of cloned channels, causing them to never be dropped. This allowed any completed threads to still ``receive" messages, which broke the terminating condition as they are never able to read them.

The solution was simple, as we could force the terminating thread to drop its receiver before exiting, disallowing any additional threads to send messages to it and continue sending messages to any of the ``live" threads.

We implemented this solution and the results are in the table below. (and also did this solution for the updated run of DPA* with the expensive heuristic and HDA*).

\begin{table}[htb]
  \centering
\begin{tabular}{|c|c|}
  \hline
  Thread Count & Run Time (Seconds) \\
  \hline
  1 & 0.050088 \\ 
  \hline
  2 & 0.086390 \\ 
    \hline
  4 & 0.077879 \\ 
    \hline
  8 & 0.089861 \\ 
    \hline
  16 & 0.21919 \\ 
  \hline
\end{tabular}
\caption{Final DPA* results with Manhattan heuristic on a 1,000,000 node search space}
\label{table:dpa_man_updated}
\end{table}

\subsubsection{Expensive Heuristic (Initial Run)}

\begin{table}[htb]
  \centering
\begin{tabular}{|c|c|}
  \hline
  Thread Count & Run Time (Seconds) \\
  \hline
  1 & 1.8387 \\ 
  \hline
  2 & 2.5358 \\ 
    \hline
  4 & 1.5934 \\ 
    \hline
\end{tabular}
\caption{Initial DPA* results with Expensive heuristic on a 10,000 node search space}
\label{table:dpa_exp}
\end{table}

Similar to the initial run of the Euclidean (cheap) heuristic above you may immediately notice the absence of the 8 and 16 threads. This is due to the same reasons discussed above.

Looking at the results from the table above, interesting results are shown to give benefit to more cores. Over the average of hundreds of runs, 1 thread with DPA* on the expensive heuristic gave a roughly $~1.8$ second runtime. 

If we move to 2 threads, keeping all other variables the same, the speed gets worse at $~2.5$ seconds. We theorize this to be because of constant overhead in having multiple threads and performing waits on some variables. 

Interestingly, if we move to 4 thread we consistently get the best performance clocking in at $~1.6$ seconds. So moving from 1 to 4 threads does show a noticeable and consistent performance increase. Likely because the 2 new threads didn't add any new overhead but just helped more evenly distribute the workload of heuristic calculation and node expansion.

\subsubsection{Expensive Heuristic (Updated Run)}

Using the update described in \ref{update} Cheap Heuristic (Updated Run) we were able to run with 8 and 16 threads here. Below is the result of that run.

\begin{table}[htb]
  \centering
\begin{tabular}{|c|c|}
  \hline
  Thread Count & Run Time (Seconds) \\
  \hline
  1 & 1.6721 \\ 
  \hline
  2 & 2.5114 \\ 
    \hline
  4 & 1.5800 \\ 
    \hline
  8 & 0.80586 \\ 
    \hline
  16 & 0.43366 \\ 
  \hline
\end{tabular}
\caption{Final DPA* results with Expensive heuristic on a 10,000 node search space}
\label{table:dpa_exp}
\end{table}

From the addition of the 8 and 16 thread data points we can see that after the initial constant overhead of having multiple threads is surpassed by the increased performance of having multiple threads very good results start to be seen. The 16 thread run is 4 times as fast as the original 1 thread run with the expensive heuristic. 

\subsection{HDA*}

\subsubsection{Background}

Previous experimentation on HDA* noted its benefit over DPA* with the idea of state ownership amongst certain processors. Testing was done using HPC clusters of various sizes packing many messages into 16+ CPU machines and less into smaller machines, it was noted that performance was highly dependent on the number of physical CPU cores used and the individual speed of said cores. \cite{survey_p_astar} Again, our testing will be done on various search spaces with 4 physical CPU cores.

\subsubsection{Cheap Heuristic (Initial Run)}

From the table below we can see that HDA* is more performant than DPA* at all thread levels - this holds with our initial hypothesis as well as the results of related work. \cite{survey_p_astar} We can also see that like DPA* with the Euclidean heuristic there is an increase in runtime when adding threads (obviously not ideal) but this time the not as large. We see that moving from 1 to 4 threads is a 1.54x increase, where in DPA* it was a 2.71x increase which is much worse. From this we can see the overhead on HDA* is lower. 

\begin{table}[htb]
  \centering
\begin{tabular}{|c|c|}
  \hline
  Thread Count & Run Time (Seconds) \\
  \hline
  1 & 0.088057 \\ 
  \hline
  2 & 0.1103 \\ 
    \hline
  4 & 0.13609 \\ 
    \hline
\end{tabular}
\caption{Initial HDA* results with Euclidean heuristic on a 1,000,000 node search space}
\label{table:hda_euc}
\end{table}

The below table provides the same test as above (HDA*, 1,000,000 node graph) but using the Manhattan heuristic as it provides a more accurate representation of the movement.

\begin{table}[htb]
  \centering
\begin{tabular}{|c|c|}
  \hline
  Thread Count & Run Time (Seconds) \\
  \hline
  1 & 0.042079 \\ 
  \hline
  2 & 0.061875 \\ 
    \hline
  4 & 0.065989 \\ 
    \hline
  8 & 0.094115 \\ 
    \hline
\end{tabular}
\caption{Initial HDA* results with Manhattan heuristic on a 1,000,000 node search space}
\label{table:hda_man}
\end{table}

The same benefits with switching DPA* to the Manhattan heuristic are seen here. We are able to run 8 threads much more consistently on HDA*, and the runtimes are significantly improved.

\subsubsection{Cheap Heuristic (Updated Run)}

Using the update in \ref{update} Cheap Heuristic (Updated Run) the table below has the HDA* result for the 16 thread run on the Manhattan heuristic as well. It also includes changing the one \textit{Mutex} to an atomic.

\begin{table}[htb]
  \centering
\begin{tabular}{|c|c|}
  \hline
  Thread Count & Run Time (Seconds) \\
  \hline
  1 & 0.042384 \\ 
  \hline
  2 & 0.044140 \\ 
    \hline
  4 & 0.041139 \\ 
    \hline
  8 & 0.074064 \\ 
    \hline
  16 & 0.13072 \\ 
  \hline
\end{tabular}
\caption{Final HDA* results with Manhattan heuristic on a 1,000,000 node search space}
\label{table:hda_man}
\end{table}

The additional data point above doesn't add too much insight as it follows the trend with HDA* on the Manhattan heuristic from the previous section (where adding more threads slightly decreased performance) however it validates that the program runs properly. However the overall inclusion of atomic showed significant performance gains similar to its inclusion on DPA* (see No Locks section on DPA*).

\subsubsection{Expensive Heuristic (Initial Run)}

Running HDA* with the Expensive heuristic gave results that consistently improved with additional threads as you can see in the table below.

\begin{table}[htb]
  \centering
\begin{tabular}{|c|c|}
  \hline
  Thread Count & Run Time (Seconds) \\
  \hline
  1 & 1.6848 \\ 
  \hline
  2 & 1.6046 \\ 
    \hline
  4 & 1.1367 \\ 
    \hline
\end{tabular}
\caption{Initial HDA* results with Expensive heuristic on a 10,000 node search space}
\label{table:hda_exp}
\end{table}

If we look back at DPA* with the expensive heuristic runtime was worsened when the second thread was added (because of higher overhead) but with HDA* we see a decreased runtime on the addition of the second thread and on the addition of the third/fourth threads.

The runtime between 1 and 2 threads was an improvement of $~5\%$. And going from 2 to 4 threads yielded an improvement of $~41\%$.

\subsubsection{Expensive Heuristic (Final Run)}

Using the updates of removing the lock in favor of a compare and set loop and removing the deadlock case for larger threads the below table is the re-run of the Expensive Heuristic on HDA*. You can note how much the performance increased from the directly proceeding table.

\begin{table}[htb]
  \centering
\begin{tabular}{|c|c|}
  \hline
  Thread Count & Run Time (Seconds) \\
  \hline
  1 & 1.6226 \\ 
  \hline
  2 & 1.4744 \\ 
    \hline
  4 & 1.0803 \\ 
    \hline
  8 & 0.71974 \\ 
    \hline
  16 & 0.47127 \\ 
  \hline
\end{tabular}
\caption{Final HDA* results with Expensive heuristic on a 10,000 node search space}
\label{table:hda_exp}
\end{table}

When HDA* on the Expensive heuristic gets to 16 threads the runtime is 4x better than the single threaded implementation. And comparing to the initial run of the HDA* Expensive heuristic better performance at each thread count resulted from the removal of the lock. 

\subsection{MPA*}

Throughout our research we quickly began to realize there was not much majorly adopted GPU support within our language of choice Rust.  While there are some self-made libraries out there it was super hard to understand and did not have any official support.  After deliberation with the team we decided to still attempt to write out code in Rust for the GA* algorithm but just test it on a CPU.  This would of course greatly impact our results since the algorithm is meant to be run on a GPU but the lack of overall GPU support with a fairly new language, in Rust, was a roadblock that made GPU support outside the scope of this research. When we began implementing GA* using CPU cores instead, we realized that we didn't allow ourselves enough time to implement it. The details of GA* took far longer to grasp, and only parts of the algorithm are actually parallelized rather than being done completely in parallel. We ultimately found a repository with the algorithm implemented in CUDA, but the code was incredibly difficult to parse. Ultimately, if we'd made GA* our main focus from the start we might've been able to fully implement it, but it was too complex for what we scoped.

\section{Conclusion}

In this project we introduced Rust implementations for various version of parallel A* including KPBFS, DPA*, and HDA*. KPBFS, being the most simple solution, had lots of synchronization overhead and thus had the overall worst performance and in cheap heuristic tests suffered the most contention when adding additional threads. DPA* was a significant jump over KPBFS - it reduced overhead and overall performance increased drastically. HDA* provided a similar jump over DPA* with the introduction of hashing the search space. With cheap heuristics performance remained flat with the addition of more threads (with DPA* and HDA*) or increased (KPBFS). Thought this was expected as shown in previous work. \cite{kpbfs}\cite{cen_decen} When an expensive heuristic was used all algorithms saw great performance increases with the addition of more threads.

One interesting finding made was depending on how accurately the heuristic modeled the movement of the algorithm affected performance. Since A* is just a Breadth-First Search with an additional heuristic if you have inadmissible (over-estimation) you will just get horrible results and runtimes and if you have heuristics that under-estimate too much the algorithm's runtime just turns into that of the size of the search space (like BFS). All the movements of KPBFS, DPA*, and HDA* were cardinal (or manhattan). When we switched from a euclidean heuristic (which would be an underestimation) to a manhattan heuristic (which estimates the movement most accurately) the performance roughly doubled. It is important to note that the manhattan distance heuristic isn't $100\%$ representative of the distance needed to travel because of walls in the graph (if there were no walls the results would be boring) but rather just of the movement. 

Overall, the problem of parallelizing A* still remains a very open one as evidenced by the variety of solutions and recent research and we provided implementations of common algorithms in Rust as well as some findings specific to the language and our implementations.

The repository associated with this paper is available at \href{github.com/brettfazio/parallel-astar-rust}{github.com/brettfazio/parallel-astar-rust}.

\bibliographystyle{ieeetr}
\bibliography{bib}

\begin{thebibliography}{1}

\bibitem{rust}
N.~D. Matsakis and F.~S. Klock~II, ``The rust language,'' in {\em ACM SIGAda
  Ada Letters}, vol.~34, pp.~103--104, ACM, 2014.

\bibitem{4082128}
P.~E. {Hart}, N.~J. {Nilsson}, and B.~{Raphael}, ``A formal basis for the
  heuristic determination of minimum cost paths,'' {\em IEEE Transactions on
  Systems Science and Cybernetics}, vol.~4, no.~2, pp.~100--107, 1968.

\bibitem{survey_p_astar}
A.~Fukunaga, A.~Botea, Y.~Jinnai, and A.~Kishimoto, ``A survey of parallel
  a*,'' 08 2017.

\bibitem{cen_decen}
V.~Kumar, K.~Ramesh, and V.~Rao, ``Parallel best-first search of state-space
  graphs: A summary of results,'' vol.~88, pp.~122--127, 01 1988.

\bibitem{k_bfs}
A.~Felner, S.~Kraus, and R.~E. Korf, ``Kbfs: K-best-first search,'' {\em Annals
  of Mathematics and Artificial Intelligence}, vol.~39, no.~1, pp.~19--39,
  2003.

\bibitem{kpbfs}
V.~Vidal, L.~Bordeaux, and Y.~Hamadi, ``Adaptive k-parallel best-first search:
  A simple but efficient algorithm for multi-core domain-independent
  planning,'' 01 2010.

\bibitem{2017arXiv170805296F}
A.~{Fukunaga}, A.~{Botea}, Y.~{Jinnai}, and A.~{Kishimoto}, ``{A Survey of
  Parallel A*},'' {\em arXiv e-prints}, p.~arXiv:1708.05296, Aug. 2017.

\bibitem{survey_p_mpastar}
Y.~Zhou and J.~Zeng, ``Massively parallel a* search on a gpu,'' in {\em
  Proceedings of the Twenty-Ninth AAAI Conference on Artificial Intelligence},
  AAAI'15, p.~1248–1254, AAAI Press, 2015.

\end{thebibliography}











\end{document}